\documentclass[showpacs,aps,twocolumn]{revtex4}
\usepackage{epsfig}
\usepackage{graphicx}
\usepackage{amsmath,amssymb,amsfonts}
\usepackage{array}
\usepackage{url}
\usepackage{hyperref}
\usepackage{subfig}
\usepackage{multirow}
\usepackage{float}
\usepackage{lineno}
\usepackage{xspace}
\usepackage[usenames,dvipsnames]{color}
\definecolor{kugray5}{RGB}{224,224,224}
\usepackage{adjustbox}
\usepackage{booktabs}
\usepackage{multirow}
\usepackage{rotating,tabularx}
\usepackage{lscape}
\newcommand{\Ksall}{$K^{0}_{S}$, $\Lambda+\bar{\Lambda}$, $\Xi^{-}+\bar{\Xi}^{+}$ and $\Omega^{-}+\bar{\Omega}^{+}$\xspace}

\newcommand{\bef}{\begin{figure}}
\newcommand{\eef}{\end{figure}}
\newcommand{\bc}{\begin{center}}
\newcommand{\ec}{\end{center}}

\newcommand{\be}{\begin{equation}}
\newcommand{\ee}{\end{equation}}
\newcommand{\bea}{\begin{eqnarray}}
\newcommand{\eea}{\end{eqnarray}}

\begin{document}
\title{Multiplicity Dependence of Non-extensive Parameters for Strange and Multi-Strange Particles in Proton-Proton Collisions at $\sqrt{s}= 7$ TeV at the LHC}
\author{Arvind Khuntia}
\author{Sushanta Tripathy}
\author{Raghunath Sahoo\footnote{Corresponding author: $Raghunath.Sahoo@cern.ch$}}
%\email{Raghunath.Sahoo@cern.ch}
\affiliation{Discipline of Physics, School of Basic Sciences,\\ Indian Institute of Technology Indore, Indore-453552, India.}
\author{Jean Cleymans}
\affiliation{UCT-CERN Research Centre and Department of Physics, University of Cape Town, Rondebosch 7701, South Africa}

\begin{abstract}
  The transverse momentum ($p_{T}$) spectra in proton-proton collisions at $\sqrt{s}$ = 7 TeV, measured by the ALICE experiment at the LHC are analyzed with a thermodynamically consistent Tsallis distribution. The information about the freeze-out surface in terms of freeze-out volume, temperature and the non-extensivity parameter, $q$, for \Ksall are extracted by fitting the $p_{T}$ spectra with Tsallis distribution function. The freeze-out parameters of these particles are studied as a function of charged particle multiplicity density ($dN_{ch}/d\eta$). In addition, we also study these parameters as a function of particle mass to see any possible mass ordering. The strange and multi-strange particles show mass ordering in volume, temperature, non-extensive parameter and also a strong dependence on multiplicity classes. It is observed that with increase in particle multiplicity, the non-extensivity parameter, $q$ decreases, which indicates the tendency of the produced system towards thermodynamic equilibration. The increase in strange particle multiplicity is observed to be due to the increase of temperature and may not be due to the size of the freeze-out volume.
\end{abstract}
\date{\today}
\maketitle

\section{Introduction}
\label{intro}
 High-energy heavy-ion collisions at RHIC and LHC provide a unique opportunity to study nuclear matter under extreme conditions {\it i.e.} at high temperature and/or energy density. The production of strange hadrons in high-energy collisions provides a unique tool to study the strongly interacting matter, which is governed by partonic degrees of freedom and is described by Quantum Chromo Dynamics (QCD). Considering the constituent quarks, in hadronic and nuclear collisions, only the light quarks, namely $u$ and $d$, participate as the colliding quanta of the system. The strange quarks are produced via hard processes, by annihilation
of light quark-antiquark pairs ($q\bar{q}~\to~s\bar{s}$) and gluon fusion ($gg~\to~s\bar{s}$) \cite{Rafelski:1982pu} during the early stages of the collisions. The enhancement of the strange hadron production thus serves as one of the signatures of Quark-Gluon Plasma (QGP) production, where quarks and gluons are no longer confined within the hadronic domain, rather the partonic degrees of freedom become important in nuclear dimensions.
%The high-multiplicity $p+p$ collisions at the LHC have become a highly interesting subject of studies because one observes the onset of higher radial collective flow \cite{Adam:2016emw}, which often is considered as one of the signatures of QGP. 
The strange particle enhancement, when studied as a function of charged particle multiplicity in $p+p$ collisions at $\sqrt s=$7 TeV shows an enhanced production of strangeness compared to the non-strange hadrons \cite{Adam:2016emw}. This heavy-ion like observation of strangeness enhancement, which is not understood from the available theoretical models, makes the high-multiplicity events at the LHC, a very interesting subject of studies. 
%Further, strange particle enhancement as a function of charged particle multiplicity density, which was originally considered as one of the %signatures of QGP, had so far not been observed  in $p+p$ collisions. However, recent enhancement in multiplicity dependence study of $p_T$ %integrated yield of strange and multi-strange relative to the pions in pp at $\sqrt s=$7 TeV \cite{Adam:2016emw} increases significantly which %resembles the behavior to $p+Pb$ collisions with the same multiplicity densities at slightly smaller center of mass energy. 
Again the hardening of the $p_T$-spectra for high-multiplicity $p+p$ collisions at $\sqrt s=$7 TeV indicates collective expansion of the system in the final state. The mass ordering of the freeze-out parameters has been observed at LHC \cite{Thakur:2016boy}, however the multiplicity dependence study of the freeze-out parameters has not been done and it will be more interesting as high-multiplicity $p+p$ events behave similar to p-Pb with the same multiplicity densities with lesser center-of-mass energy \cite{Abelev:2013haa}. Hence, the multiplicity dependence analysis in $p+p$ collisions becomes interesting in view of higher radial flow and other QGP signatures, as seen in A+A collisions. To understand the event dynamics in these high-multiplicity $p+p$ collisions, from the thermodynamic viewpoints, we analyze the spectra of strange and multi-strange particles in the domain of non-extensive statistics, which becomes inevitable at the high-$p_T$ reach in high-energy collisions at the LHC. It has also been proposed recently that like A+A collisions, the $p+p$ collisions at higher collision energies are better described by a grand canonical ensemble, making the latter events more interesting \cite{Das:2016muc}.
  
  \bef[H]
\begin{center}
\includegraphics[scale=0.45]{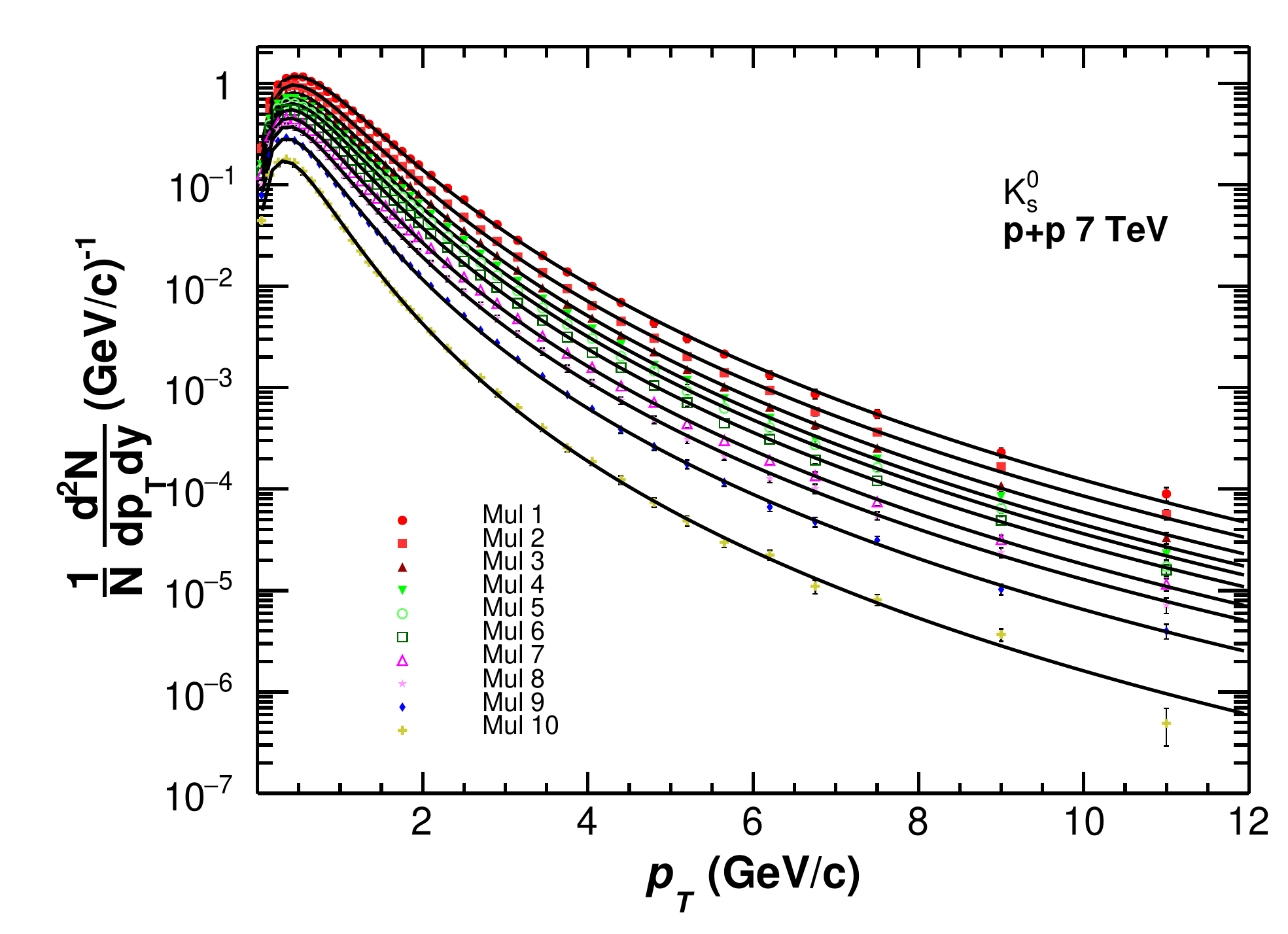}
\newline
\caption{(Color online) Fitting of $p_T$-spectra of $K^{0}_{S}$ \cite{Adam:2016emw} with Tsallis distribution for $p+p$ collisions 
at $\sqrt{\mathrm{s}}$= 7 TeV using Eq.~\ref{eq6} for various multiplicity classes as given in Table~\ref{table:mult_info}.}
\label{ffit:Tsallis:Ks}
\end{center}
\eef

\bef[H]
\begin{center}
\includegraphics[scale=0.45]{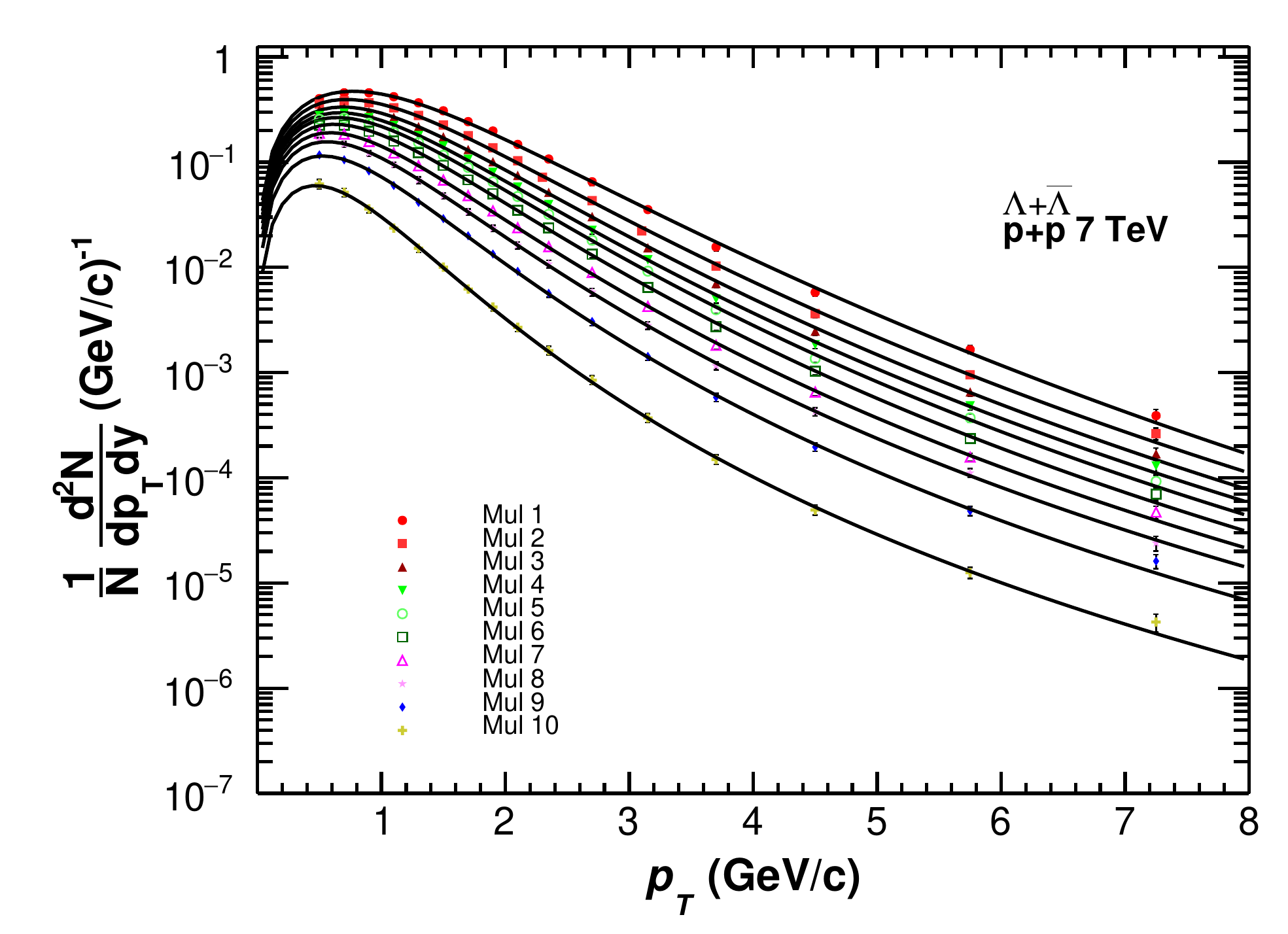}
\newline
\caption{(Color online) Fitting of $p_T$-spectra of $\Lambda+\bar{\Lambda}$ \cite{Adam:2016emw} with Tsallis distribution for $p+p$ collisions 
at $\sqrt{\mathrm{s}}$= 7 TeV using Eq.~\ref{eq6} for various multiplicity classes as given in Table~\ref{table:mult_info}.}
\label{ffit:Tsallis:L}
\end{center}
\eef

 At higher center-of-mass energies the probability of multi-partonic interaction increases, where there is more than one hard scattering, which leads to high-multiplicity $p+p$ collisions. Due to the high multiplicities produced in $p+p$ collisions, one can use the the statistical models  to describe the particle production mechanism. Such a statistical description of transverse momentum $(p_{T})$ of final state particles produced in high-energy collisions has been proposed to follow a thermalized Boltzmann type of distribution as given by  \cite{Hagedorn:1965st}

         \begin{eqnarray}
         \label{eq1}
         E\frac{d^3\sigma}{d^3p}& \simeq C \exp\left(-\frac{p_T}{T_{exp}}\right).
         \end{eqnarray}
         
However, the identified particle spectra at RHIC and LHC do not follow Boltzmann-Gibbs distribution due to the possible QCD contributions at high-$p_{T}$. To account for this effect a power-law in $p_{T}$ has been proposed \cite{CM,CM1,UA1}, which empirically accounts for the possible QCD contributions. Hagedorn proposed a combination of both aspects, which describes the experimental data over a wide  $p_{T}$ range  \cite{Hagedorn:1983wk} and is given by 
  
\begin{eqnarray}
  E\frac{d^3\sigma}{d^3p}& = &C\left( 1 + \frac{p_T}{p_0}\right)^{-n}
\nonumber\\
 & \longrightarrow&
  \left\{
 \begin{array}{l}
  \exp\left(-\frac{n p_T}{p_0}\right)\quad \, \, \, {\rm for}\ p_T \to 0, \smallskip\\
  \left(\frac{p_0}{p_T}\right)^{n}\qquad \qquad{\rm for}\ p_T \to \infty,
 \end{array}
 \right .
 \label{eq2}
\end{eqnarray}

where $C$, $p_0$, and $n$ are fitting parameters. This function behaves like an exponential  for small $p_{T}$ and  like a power-law  for large $p_{T}$ values. A finite degree of deviation from the equilibrium statistical description of identified particle $p_{T}$ spectra has already been observed by experiments at RHIC \cite{star-prc75,phenix-prc83} and LHC \cite{alice1,alice2,alice3,cms}.  For a thermalized system the $\langle p_{T} \rangle$ is associated with the temperature of the hadronizing matter, but for systems  which are far from thermal equilibrium one fails to make such a connection. In the latter systems, either the temperature fluctuates event-by-event or within the same event \cite{Bhattacharyya:2015nwa}. This creates room for a possible description of the $p_{T}$ spectra in high-energy hadronic and nuclear collisions, using the non-extensive Tsallis statistics \cite{Tsallis:1987eu,Tsallis:2008mc,book}. A thermodynamically consistent  non-extensive distribution function is given by \cite{Cleymans:2011in} 
    \begin{equation}
\label{eq3}
f(m_T) =  C_q \left[1+{(q-1)}{\frac{m_T}{T}}\right]^{-\frac{1}{q-1}} .
\end{equation}
  Here, $m_{\rm T}$ is the transverse mass and $q$ is called the non-extensive parameter- a measure of the degree of deviation from equilibrium. Eqs. \ref{eq2} and \ref{eq3} are related through the following transformations for large values of $p_{T}$:
  \begin{equation}
  n= \frac{1}{q-1}, ~\mathrm{and} ~~~~ p_0 = \frac{T}{q-1}.
  \label{eq4}
  \end{equation} 

\par
 In the limit $q \rightarrow 1$, one recovers the standard Boltzmann-Gibbs distribution (Eq. \ref{eq1}) from the Tsallis distribution  (Eq. \ref{eq3}). 
      
Tsallis non-extensive statistics is used widely to explain the particle spectra in high-energy collisions \cite{Bhattacharyya:2015nwa,Bhattacharyya:2015hya,Zheng:2015gaa,Tang:2008ud,De:2014dna} starting from elementary 
$e^++e^-$, hadronic and heavy-ion collisions \cite{e+e-,R1,R2,R3,ijmpa,plbwilk,marques,STAR,PHENIX1,PHENIX2,ALICE_charged,ALICE_piplus,CMS1,CMS2,ATLAS,ALICE_PbPb}. In the framework of non-extensive statistics, the nuclear modification factor has been successfully described for the heavy-flavor particles at the RHIC and the LHC using Boltzmann Transport Equation with a relaxation time approximation \cite{Tripathy:2016hlg}. The criticality in the non-extensive $q$-parameter is also shown for speed of sound in a hadron resonance gas using non-extensive statistics \cite{Khuntia:2016ikm}. A comprehensive analysis of $\pi^-$ and heavy quarkonium states has recently been done in Ref. \cite{Grigoryan:2017gcg,Parvan:2016rln}.
 
The paper is organized as follows. In section 2,  we present  the thermodynamically consistent Tsallis distribution function, which is used to describe the particle spectra. Then we discuss the results in view of the non-extensive statistical description of the strange and multi-strange particle spectra as a function of charged particle multiplicity density and particle mass. Finally, in section 3 we present a summary of our results. 

%%%%%%%%%%%%%%%%%%%%%%%%%%%%%%%%%%%%%%%%%%%%%%%%%%%%%%%%%%%%%%%%%%%%%%%%%%%%%
\begin{widetext}
\begin{table*}[htbp]
\caption[p]{Number of mean charged particle multiplicity density corresponding to different event classes \cite{Adam:2016emw}.}
\label{table:mult_info}
%\centering
\begin{adjustbox}{max width=\textwidth}
\begin{tabular}{c|c|c|c|c|c|c|c|c|c|c|c|}
\hline
\multicolumn{2}{|c|}{${\bf Class name}$}&Mul1&Mul2&Mul3&Mul4&Mul5&Mul6&Mul7&Mul8&Mul9&Mul10\\
\hline

\multicolumn{2}{|c|}{ $  \bf \big<{\frac{dN_{ch}}{d\eta} } \big>$} &21.3$\pm$0.6&16.5$\pm$0.5&13.5$\pm$0.4&11.5$\pm$0.3&10.1$\pm$0.3&8.45$\pm$0.25&6.72$\pm$0.21&5.40$\pm$0.17&3.90$\pm$0.14&2.26$\pm$0.12\\
\hline
\end{tabular}
\end{adjustbox}
 \end{table*}
 \end{widetext}

%%%%%%%%%%%%%%%%%%%%%%%%%%%%%%%%%%%%%%%%%%%%%%%%%%%%%%%%%%%%%%%%%%%%%%%%%%%%%

\bef[H]
\begin{center}
\includegraphics[scale=0.45]{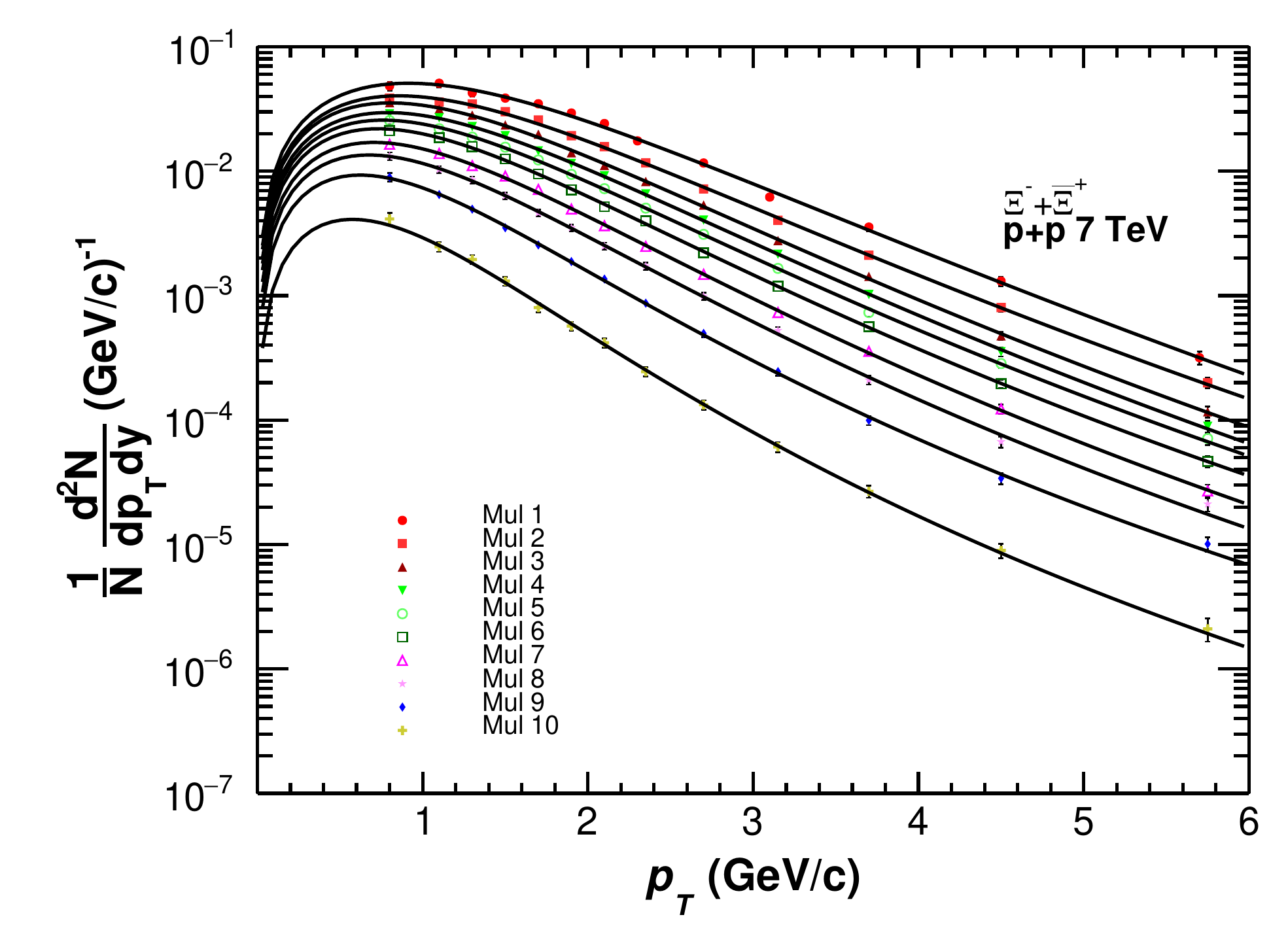}
\newline
\caption{(Color online) Fitting of $p_T$-spectra of $\Xi^{-}+\bar{\Xi}^{+}$ \cite{Adam:2016emw} with Tsallis distribution for $p+p$ collisions 
at $\sqrt{\mathrm{s}}$= 7 TeV using Eq.~\ref{eq6} for various multiplicity classes as given in Table~\ref{table:mult_info}.}
\label{ffit:Tsallis:Cascade}
\end{center}
\eef

\bef[H]
\begin{center}
\includegraphics[scale=0.45]{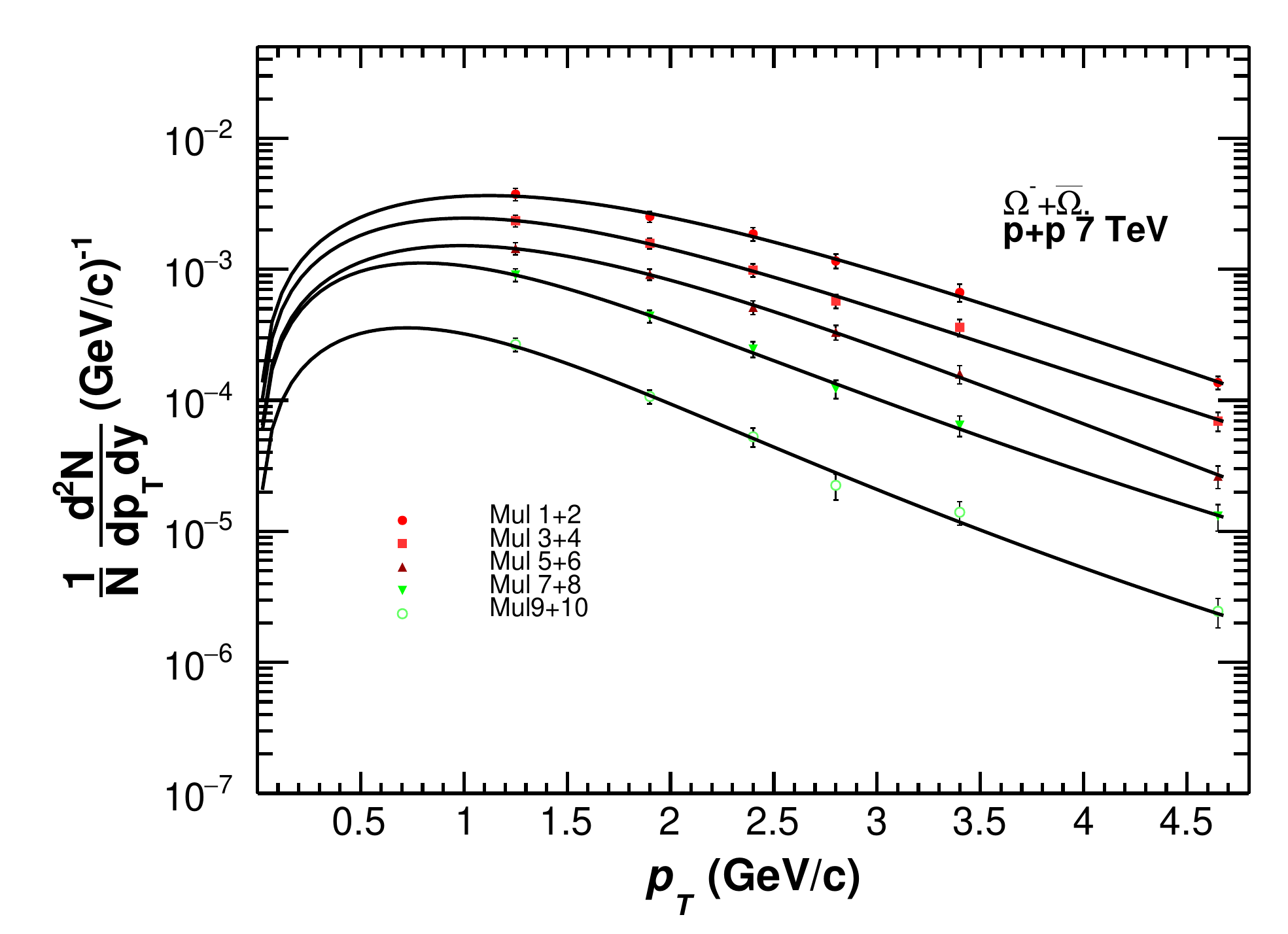}
\newline
\caption{(Color online) Fitting of $p_T$-spectra of $\Omega^{-}+\bar{\Omega}^{+}$  \cite{Adam:2016emw} with Tsallis distribution for $p+p$ collisions 
at $\sqrt{\mathrm{s}}$= 7 TeV using Eq.~\ref{eq6} for various multiplicity classes as given in Table~\ref{table:mult_info}.} 
\label{ffit:Tsallis:O}
\end{center}
\eef

%%%%%%%%%%%%%%%%%%%%%%%%%%%%%%%%%%%%%%%%%%%%%%%%%%%%%%%%
\begin{widetext}
%\begin{table*}[p]
\begin{table*}[htbp]
\caption{The extracted Tsallis parameters as well as the $\chi^2$/ndf  for all the multiplicity classes. For $\Omega^{-}+\bar{\Omega}^{+}$, multiplicity classes are combined to deal with the statistics \cite{Adam:2016emw}.}
\label{table:parameters}
\newcommand{\tabincell}
\centering
%\begin{center}
\begin{adjustbox}{max width=\textwidth}
%\begin{sidewaystable}[p!]
\begin{tabular}{|c|c|c|c|c|c|c|c|c|c|c|c|}\toprule 

\hline
\multicolumn{2}{|l|}{${\bf Particles}$}&\multicolumn{10}{c|}{\bf Multiplicity class} \\
\cline{3-12}
\multicolumn{2}{|c|}{} &{\bf Mul1} & {\bf Mul2} & {\bf Mul3} & {\bf Mul4}& {\bf Mul5} &{\bf Mul6} & {\bf Mul7} & {\bf Mul8} & {\bf Mul9}&{\bf Mul10}\\
\hline
\multirow{3}{*}{$\bf{K^{0}_{S}}$}& T (GeV) &0.152$\pm$0.001&0.137$\pm$0.001&0.131$\pm$0.004&0.124$\pm$0.003&0.119$\pm$0.005&0.111$\pm$0.004&0.103$\pm$0.004&0.095$\pm$0.002&0.085$\pm$0.003&0.068$\pm$0.003\\
\cline{2-12}
                   
                    & q &1.141$\pm$0.001 & 1.144$\pm$0.001 & 1.144$\pm$0.002&1.145$\pm$0.002 &1.146$\pm$0.003&1.148$\pm$0.002&1.148$\pm$0.002&1.150$\pm$0.002&1.150$\pm$0.002&1.147$\pm$0.002\\
                    \cline{2-12}
                    & $\chi^2$/ndf& 0.275&0.429&0.235&0.323&0.309&0.286&0.484&0.384&0.321&0.494\\  
\hline
%%%%%%%%%%%%%%%%22222222222%%%%%%%%%%%%%%%%%%%%%%%%%
\multirow{3}{*}{$\bf{\Lambda+\bar{\Lambda}}$}&T (GeV)&0.245$\pm$ 0.0&0.201$\pm$ 0.0&0.179$\pm$0.0&0.159$\pm$0.0&0.146$\pm$0.0&0.128$\pm$0.0&0.102$\pm$0.0&0.082$\pm$0.0&0.056$\pm$0.0&0.010$\pm$0.0\\
                   
                   \cline{2-12} 
                    & q & 1.086$\pm$0.006& 1.097$\pm$0.004& 1.101$\pm$0.007&1.106$\pm$0.004 &1.108$\pm$0.004&1.111$\pm$0.001&1.118$\pm$0.003&1.123$\pm$0.002&1.128$\pm$0.004&1.139$\pm$.001\\
                    \cline{2-12}
                    & $\chi^2$/ndf& 0.554&0.543&0.281&0.311&0.272&0.307&0.201&0.160&0.312&0.248\\

\hline
%%%%%%%%%%%%%%%%333333333333%%%%%%%%%%%%%%%%%%%%%%%%%
\multirow{3}{*}{$\bf {\Xi^{-}+\bar{\Xi}^{+}}$}&T (GeV)&0.308$\pm$0.0&0.260$\pm$0.0&0.224$\pm$0.0&0.212$\pm$0.0&0.186$\pm$0.0&0.164$\pm$0.0&0.147$\pm$0.0&0.122$\pm$0.0&0.074$\pm$0.0 &0.045$\pm$0.001\\
                   
                   \cline{2-12} 
                    & q &1.069$\pm$0.015 & 1.081$\pm$0.005&1.086$\pm$0.005 &1.088$\pm$0.004 &1.096$\pm$0.003&1.100$\pm$0.003&1.101$\pm$0.003&1.108$\pm$0.002&1.121$\pm$0.002&1.122$\pm$0.002\\
                  \cline{2-12}
                    & $\chi^2$/ndf& 0.837&0.458&0.350&0.133&0.168&0.232&0.237&0.543&0.313&0.369\\

\hline 
\multicolumn{2}{c|}{}&\multicolumn{2}{c|}{\bf Mul[1+2]}&\multicolumn{2}{c|}{\bf Mul[3+4]}&\multicolumn{2}{c|}{\bf Mul[5+6]}&\multicolumn{2}{c|}{\bf Mul[7+8]}&\multicolumn{2}{c|}{\bf Mul[9+10]} \\
\hline
\multirow{3}{*}{$\bf{\Omega^{-}+\bar{\Omega}^{+}}$}& T (GeV) &\multicolumn{2}{c|}{0.422$\pm$0.0}&\multicolumn{2}{c|}{0.310$\pm$0.0}&\multicolumn{2}{c|}{0.330$\pm$0.001}&\multicolumn{2}{c|}{0.104$\pm$0.0}&\multicolumn{2}{c|}{0.052$\pm$0.0}\\
 \cline{2-12} 
  & q &\multicolumn{2}{c|}{1.035$\pm$0.028} & \multicolumn{2}{c|}{1.066$\pm$0.011}&\multicolumn{2}{c|}{1.044$\pm$0.005}&\multicolumn{2}{c|}{1.117$\pm$0.015}&\multicolumn{2}{c|}{1.124$\pm$0.006} \\
 \cline{2-12}
 & $\chi^2$/ndf& \multicolumn{2}{c|}{0.297}&\multicolumn{2}{c|}{0.478}&\multicolumn{2}{c|}{0.092}&\multicolumn{2}{c|}{0.238}&\multicolumn{2}{c|}{0.652}\\
 \hline 
 \end{tabular}

%\end{sidewaystable}
\end{adjustbox}
% \end{center}
 \end{table*}
\end{widetext}
%\clearpage

%%%%%%%%%%%%%%%%%%%%%%%%%%%%%%%%%%%%%%%%%%%%%%%%%%%%%%%%

\section{Non-extensivity and Transverse momentum spectra}
\label{sec:1}
In the  sections 2 and 3, we discuss the transverse momentum spectra of identified particles \Ksall  produced in $p+p$ collisions at $\sqrt {s}$ = 7 TeV at the LHC, measured by the ALICE experiment, using a thermodynamically consistent Tsallis non-extensive statistics.

%and \cite{RefJ}
\subsection{Non-extensive statistics }
\label{sec:2}
The Tsallis distribution function at mid-rapidity, with finite chemical potential  \cite{Cleymans:2015lxa} is given by,

\begin{eqnarray}
\label{eq5}
\left.\frac{1}{p_T}\frac{d^2N}{dp_Tdy}\right|_{y=0} = \frac{gVm_T}{(2\pi)^2}
\left[1+{(q-1)}{\frac{m_T-\mu}{T}}\right]^{-\frac{q}{q-1}}
\end{eqnarray}
 where,  $~ m_T$  is the transverse mass of a particle given by $\sqrt{p_T ^2 + m^2}$, $g$ is the degeneracy, $V$ is the system volume and $\mu$ is the chemical potential of the system. At LHC energies, where $\mu\simeq 0$, the transverse momentum distribution function~\cite{Li:2015jpa} reduces to:
\begin{eqnarray}
\label{eq6}
\left.\frac{1}{p_T}\frac{d^2N}{dp_Tdy}\right|_{y=0} = \frac{gVm_T}{(2\pi)^2}
\left[1+{(q-1)}{\frac{m_T}{T}}\right]^{-\frac{q}{q-1}}.
\end{eqnarray}
     
For extraction of  the Tsallis parameters of identified strange and multi-strange particles Eq. \ref {eq6} has been used. The degeneracy factor, $g= 2\times(2s+1)$ is taken to be 2, 8, 4 and 8 for \Ksall, respectively. Here, $s$ is the spin of the particle and the factor 2 is to take care of the antiparticles. For the case of $\Lambda$, as $\Sigma^0$ and $\Lambda$ are not experimentally distinguishable, the degeneracy factor is taken to be 8.
 \bef[H]
\begin{center}
\includegraphics[scale=0.45]{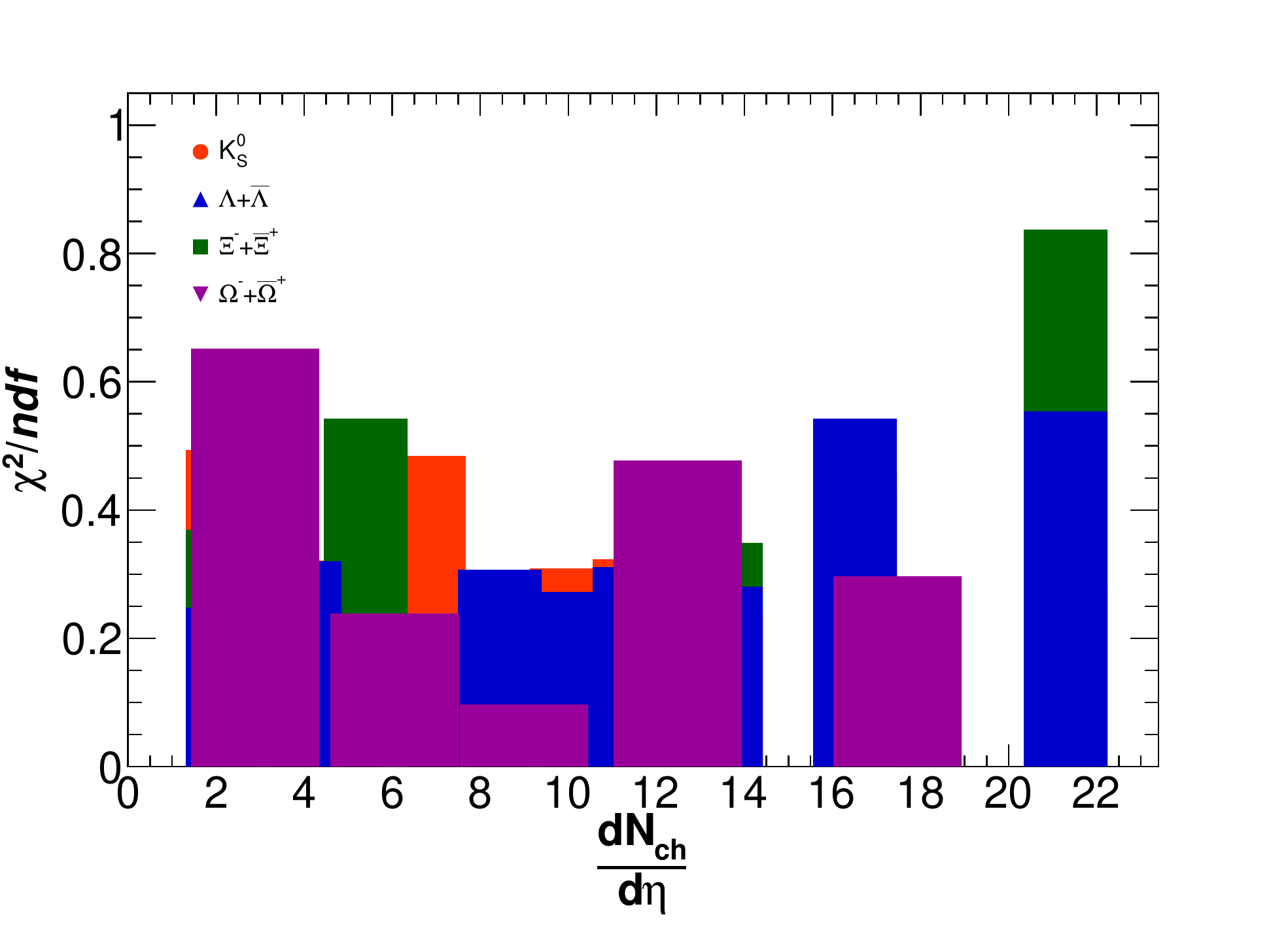}
\newline
\caption{(color online) The $\chi^2/ndf$ for all particle $p_T$-spectra fitted to the thermodynamically consistent Tsallis distribution given by Eq.~\ref{eq6}, shown as a function of charged particle multiplicity density, which signifies different class of events.}
\label{ffit:Tsallis:Chi2}
\end{center}
\eef

 %%%%%%%%%%%%%%%%%%%%%%%%%%%%%%%%%%%%%%%%%%%%%%%%
 \subsection{Results and Discussion}
The $p_{T}$-spectra for strange and multi-strange particles {\it i.e.} for $K^{0}_{S}$, $\Lambda+\bar{\Lambda}$, $\Xi^{-}+\bar{\Xi}^{+}$ and $\Omega^{-}+\bar{\Omega}^{+}$ in $p+p$ collisions at $\sqrt s=$ 7 TeV \cite{Adam:2016emw} are fitted with the thermodynamically consistent Tsallis distribution function given by Eq. \ref{eq6} for different multiplicity classes. This is shown in Figs. \ref{ffit:Tsallis:Ks}, \ref{ffit:Tsallis:L}, \ref{ffit:Tsallis:Cascade} and \ref{ffit:Tsallis:O}. The definitions of the multiplicity classes are listed in Table ~\ref{table:mult_info}. The fitting is performed using the TMinuit class available in ROOT library keeping all the parameters free. Here we follow the notion of a mass dependent differential freeze-out scenario \cite{Thakur:2016boy,Lao:2015zgd}, where particles freeze-out at different times, which correspond to different system volumes and temperatures. Then we study the thermodynamic parameters in the context of non-extensive statistics. As can be seen from Fig. \ref{ffit:Tsallis:Chi2}, for all the particle species considered in this paper, the quality of the fits given by the reduced $\chi^2$ is below one. This shows that the spectra are very well described by the non-extensive statistics. It should be noted here that the study of $\Omega$ is affected by the statistical and systematic errors because of the lower production cross-section. As the quality of the fits are very good upto high-$p_T$, with $\chi^2/ndf$ always less than 1, one doesn't need to introduce the radial flow in the theory to describe the particle spectra.

Further we have extracted the fitting parameters, namely, the radius parameter, $R$, which signifies the dimension of the system formed in these event classes; the temperature parameter; $T$ and the non-extensive parameter, $q$; which tells about the degree of deviation from a thermalized statistics, usually described by an exponential Boltzmann-Gibbs distribution function. The extracted parameters for all the particles and multiplicity classes are enlisted in Table~\ref{table:parameters}.
 
 In Fig. \ref{ffit:Tsallis:V}, the radius parameter, $R$, is shown as a function of the charged particle multiplicity density, which characterizes various event classes. Here we assume a spherical geometry of the fireball and hence the radius parameter, $R \equiv [V\frac{3}{4\pi}]^{1/3}$. However, $R$ is not necessarily related to the system size, as determined from the experimental HBT analysis, but it is related to the normalization in the statistical distribution function used to describe the particle yield/spectra \cite{Cleymans:2013rfq}. With higher multiplicity classes, $R$ is seen to decrease for all the particles with a tendency to stabilize at very high multiplicities, except for the $K^{0}_{S}$, where we found $R$ to be nearly independent of the event classes. This result is counter-intuitive as one would expect an increase in the size of the volume for higher multiplicity events. At high-multiplicity, particles emerge from a smaller volume with high decoupling temperature. This result is an important one and will be discussed again below. 

\bef[H]
\begin{center}
\includegraphics[scale=0.45]{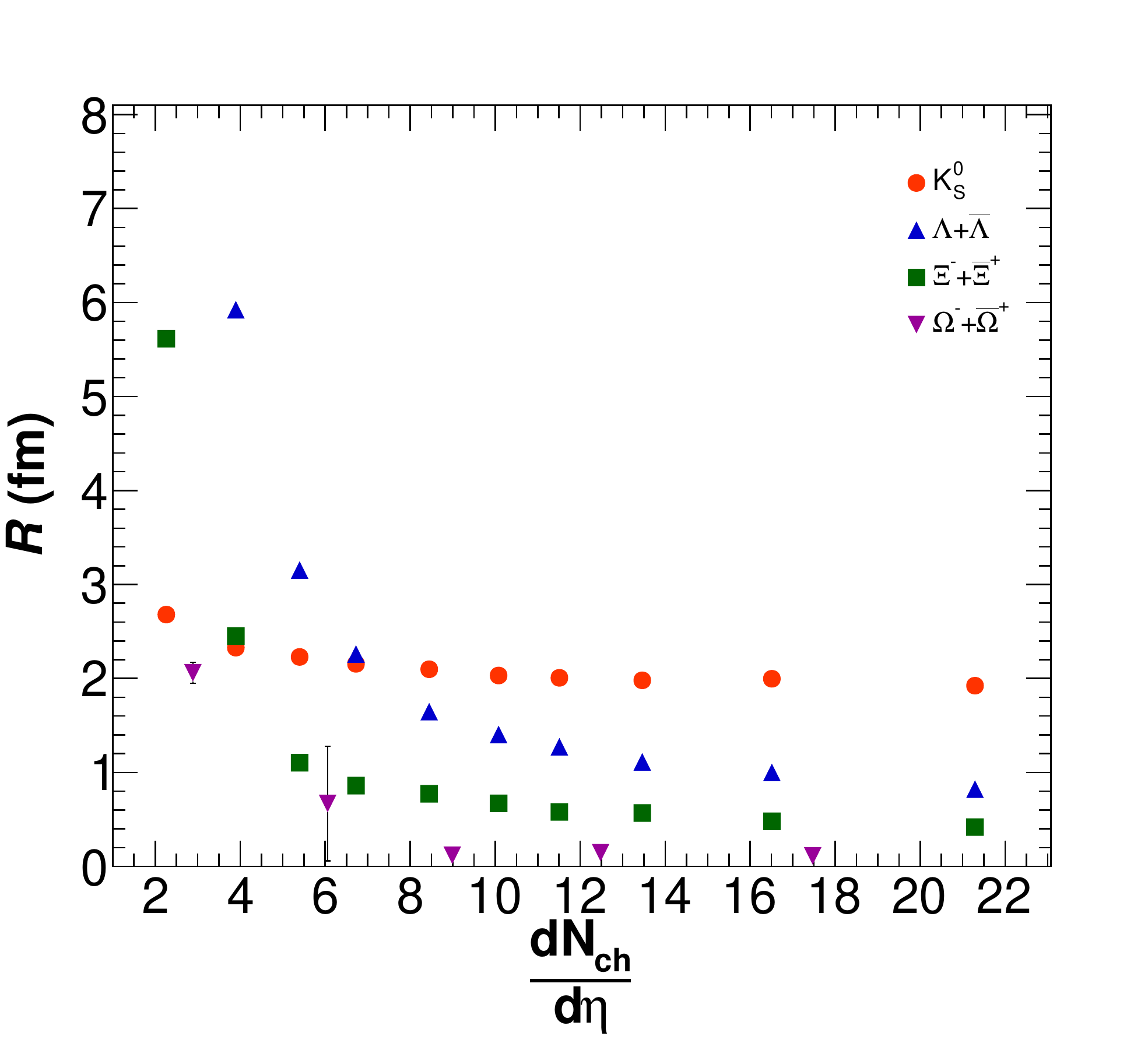}
\newline
\caption{(color online) The radius parameter, $R$, as a function of charged particle multiplicity density for different particles. $R$ is obtained from the fitting of $p_T$-spectra with Tsallis distribution by using Eq.~\ref{eq6}.}
\label{ffit:Tsallis:V}
\end{center}
\eef

In Fig. \ref{fit:Tsallis:T} we show the variation of the temperature parameter, $T$ as a function of event multiplicity. This shows a monotonic increase with increase in particle multiplicity.  For particles with higher strangeness the $T$ is seen to be higher. This indicates an early freeze-out of the multi-strange particles. This is also supported by the variation of the system size, $R$ with strangeness, {\it i.e.} the multi-strange and massive particles freeze-out early with a
 smaller system freeze-out volume. In Fig. \ref{fit:Tsallis:q}, we show the variation of the non-extensive parameter, $q$ with charged particle multiplicity. The value of $q$ decreases monotonically for higher multiplicity classes for all the particles under discussion. However, the $q$-value for $K^{0}_{S}$ is almost independent of the charged particle multiplicity density. This indicates that $K^{0}_{S}$ hardly interacts with the medium formed in the collision and thus it shows a minimal tendency for equilibration.  The fact that the $q$-values go on decreasing with multiplicity is an indicative of the tendency of the produced systems towards thermodynamic equilibrium. This goes inline with the naive expectations while understanding the microscopic view of systems approaching thermodynamic equilibrium. A similar tendency of $q$ decreasing with number of participating nucleons for Pb+Pb collisions at $\sqrt{s_{\rm NN}}$ = 2.76 TeV has been observed for the bulk part ($p_T <$ 6 GeV/c ) of the charged hadron spectra \cite{Biro:2014cka,Urmossy:2015hva}. The present studies are useful in understanding the microscopic features of degrees of equilibration and their dependencies on the number of particles in the system. The decrease in $q$ is again an important result as it indicates that
for high multiplicities the particles are produced in a hot but small volume close to a Boltzmann description. 

It should be noted here that the CMS experiment at the LHC has measured the $p_T$-spectra of $K^{0}_{S}$, 
$\Lambda+\bar{\Lambda}$, $\Xi^{-}+\bar{\Xi}^{+}$ as a function of number of tracks upto $p_T \sim$ 5 GeV/c \cite{Khachatryan:2016yru}. 
Our analysis done on these CMS data, shows the same trend for $T$ and $q$ parameters, while we observe a reverse trend of $R$ with particle multiplicity. This is presently under investigation. It could be due to the difference between ALICE using the dependence 
on $dN_{\rm ch}/d\eta$ while CMS uses the number of tracks.

\bef[H]
\begin{center}
\includegraphics[scale=0.48]{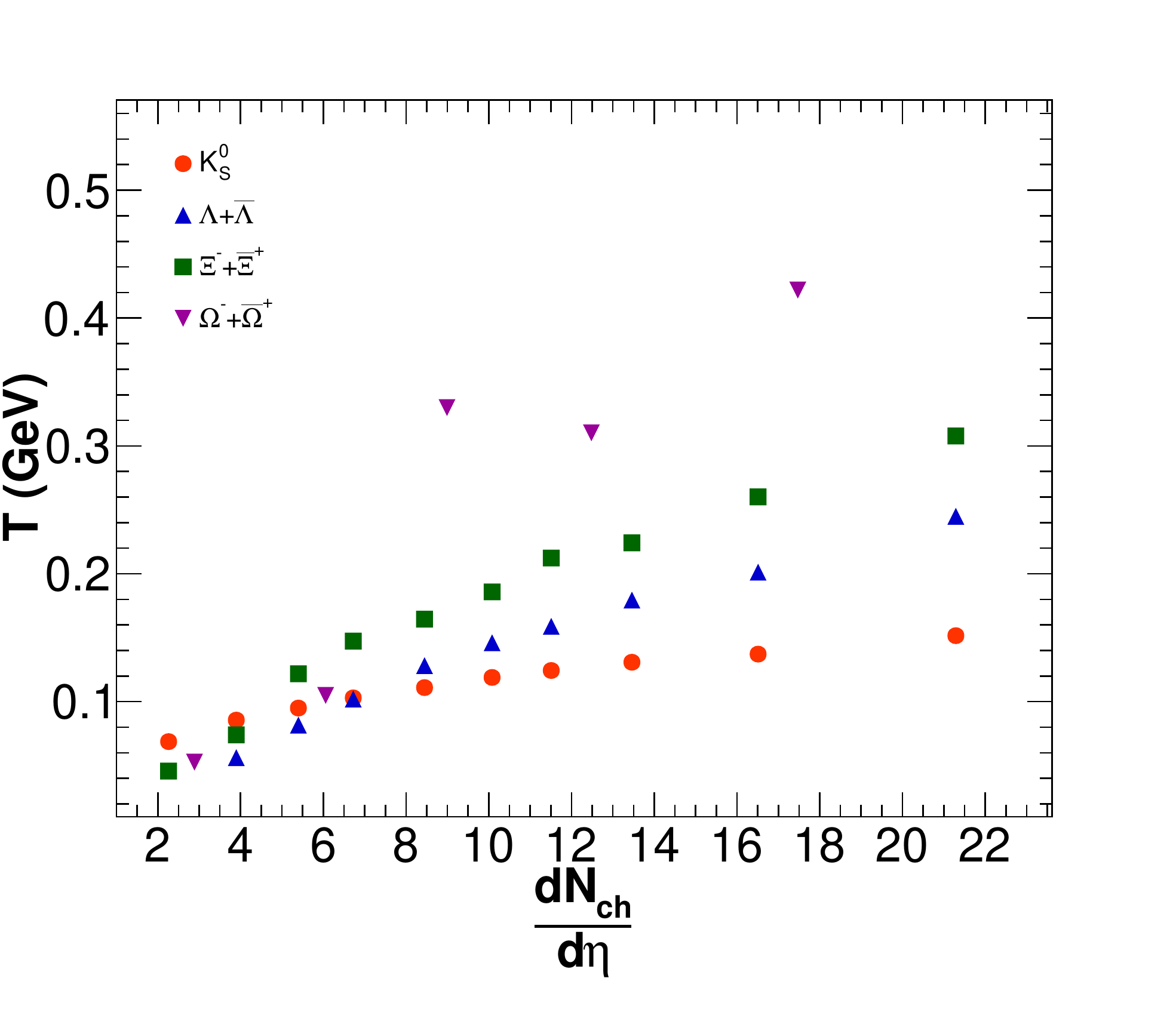}
\newline
\caption{(color online) Multiplicity dependence of $T$ for $p+p$ collisions at $\sqrt{\mathrm{s}}$= 7 TeV using Eq.~\ref{eq6} as a fitting function. }
\label{fit:Tsallis:T}
\end{center}
\eef

\bef[H]
\begin{center}
\includegraphics[scale=0.48]{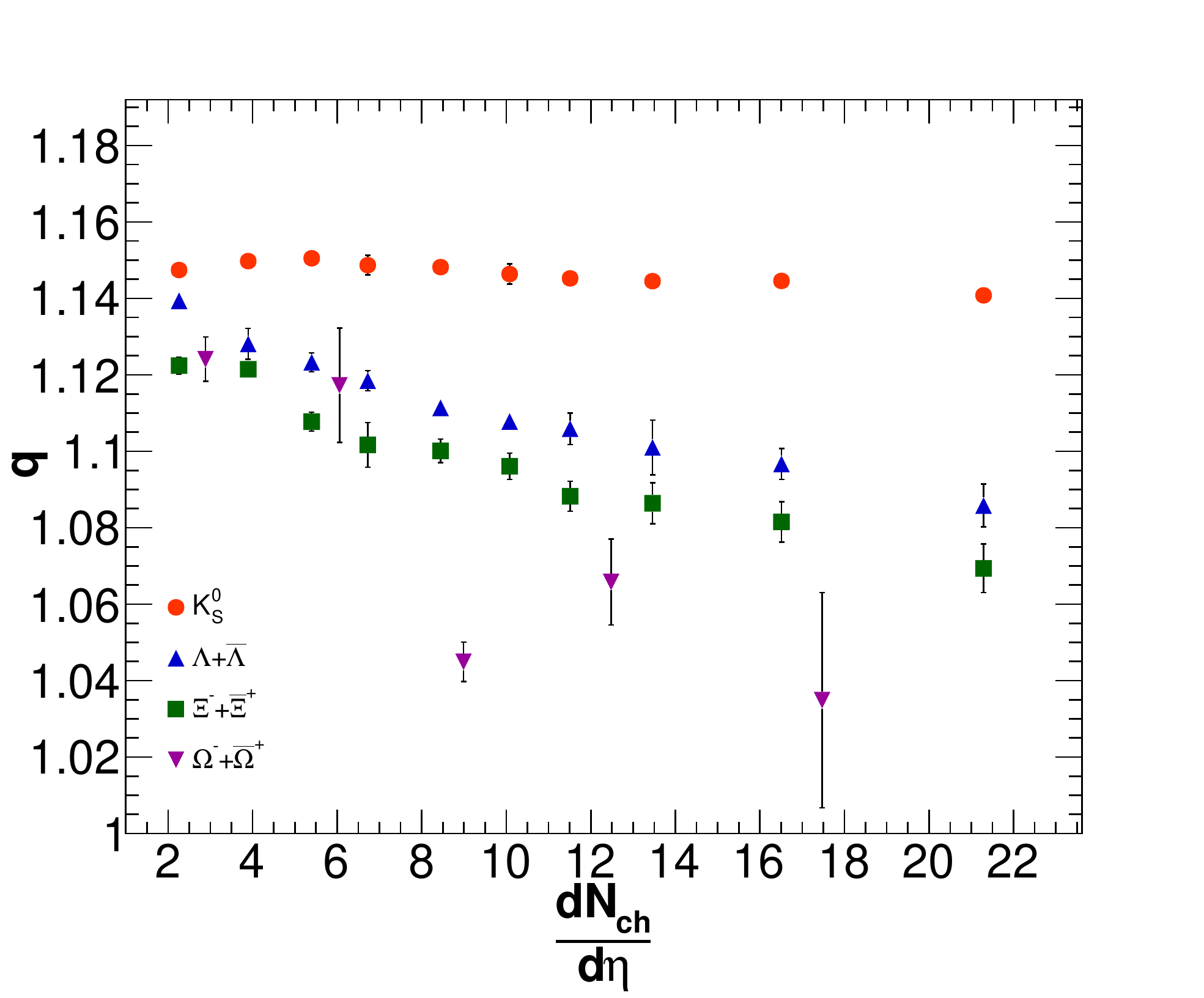}
\newline
\caption{(color online) Multiplicity dependence of the non-extensive parameter, $q$  for $p+p$ collisions at $\sqrt{\mathrm{s}}$= 7 TeV using Eq.~\ref{eq6} as a fitting function.}
\label{fit:Tsallis:q}
\end{center}
\eef

\section{Summary}
The high-multiplicity events in $p+p$ collisions at the LHC energies have become a matter of special attention to the research community, as 
it has shown heavy-ion like properties {\it e.g.}, enhancement of strange particles, which are not yet
understood from the existing theoretical models. Understanding the microscopic properties of such events are of paramount importance in order to have a complete understanding of the matter created in these collisions. In this paper, we have tried to understand these events from their thermodynamics point of view. As the $p_T$-spectra of the produced secondaries at these energies obey a combination of an empirical exponential,  {\it i.e.} thermalized Boltzmann-Gibbs type and pQCD inspired power-law, it has become customary to use a thermodynamically 
consistent Tsallis non-extensive statistics to describe these spectra. We have analysed the  multiplicity dependence of the $p_T$-spectra of strange and multi-strange particles in $p+p$ collisions at $\sqrt{s}$ = 7 TeV measured by the ALICE experiment at the LHC, using thermodynamically consistent non-extensive statistics. The extracted thermodynamic parameters like the system volume, the temperature parameter, $T$ and the non-extensive parameter, $q$ are studied as a function of charged particle multiplicity, which signifies different class 
of events. These parameters are also studied as a function of the mass of the particles. In summary,

\begin{itemize}
\item It has been shown in the present paper that the Tsallis distribution provides a very good description of the transverse momentum distributions of strange and multi-strange particles produced in $p+p$ collisions at $\sqrt{s}$ = 7 TeV without incorporating the radial flow.

\item The parameters obtained show  variations  with the multiplicity in the collision. Notably is the variation of the non-extensive parameter, $q$ which decreases towards the value 1 as the multiplicity increases, except for the $K_s^0$, which shows no clear dependence. This shows the tendency of the produced system to equilibrate with higher multiplicities. This goes inline with the expected multi-partonic interactions, which increase for
higher multiplicities in $p+p$ collisions and are thus responsible for bringing the system towards thermodynamic equilibrium.

\item The variable $T$ shows a systematic increase with multiplicity, the heaviest baryons showing the steepest increase. This is an indication of a mass hierarchy in particle freeze-out.

\item The freeze-out radius for a given particle has a tendency to remain constant at higher multiplicities. For a given multiplicity
class, the higher mass particles have lower freeze-out radii. These changes have implications for the
kinetic freeze-out conditions, where the heavy multi-strange hadrons are seen to have an earlier kinetic freeze-out,
meaning they come from a smaller volume at a higher temperature.
\end{itemize}  
These results show that the Tsallis distribution is an excellent tool to analyze high-energy $p+p$ collisions.

\section*{Acknowledgements}
ST acknowledges the financial support by DST INSPIRE program of Govt. of India. The authors gratefully acknowledge 
Dr. Sudipan De for his valuable time for reading the final manuscript.

%\appendix

%\section{Appendices}

%\clearpage

\end{document}